\documentclass[aps, pre, twocolumn, 10pt, showpacs, superscriptaddress, amsmath, amssymb, longbibliography]{revtex4-1}
\usepackage{graphicx}
\usepackage{changes}
\usepackage{mathtools}
\usepackage{subfig}
\usepackage[pdftex, colorlinks]{hyperref}
\usepackage{xcolor}
\usepackage{bbding}

\definecolor{green}{rgb}{0.0,0.26,0.15}

\newcommand{\pd}{{\mathsf{PD}}}
\newcommand{\tp}{{\mathsf{TP}}}

\hypersetup{ colorlinks, linkcolor=red, filecolor=red, urlcolor=blue,
  citecolor=blue }

\begin{document}

\title{Failure of confined granular media due to pullout of an intruder: From force networks to a system wide response}

\author{Srujal Shah} \affiliation{
School of Energy Systems, Lappeenranta-Lahti University of Technology LUT, 53851 Lappeenranta, Finland}

\author{Chao Cheng} \affiliation{
Department of Mathematical Sciences and Center for Applied Mathematics and Statistics, New Jersey Institute of Technology, Newark, New Jersey 07102, USA
}

\author{Payman Jalali} \affiliation{
School of Energy Systems, Lappeenranta-Lahti University of Technology LUT, 53851 Lappeenranta, Finland}

\author{Lou Kondic} \affiliation{
Department of Mathematical Sciences and Center for Applied Mathematics and Statistics, New Jersey Institute of Technology, Newark, New Jersey 07102, USA
}

\thanks{Corresponding author:}
\email{kondic@njit.edu}

\begin{abstract}
 We investigate computationally the pullout of a spherical intruder initially buried at the bottom of a granular column. The intruder starts to move out of the granular bed once the
pulling force reaches a critical value, leading to material failure. The failure point is found to depend on the diameter of the granular column, pointing out the importance of particle-wall interaction in determining the material response. Discrete element simulations show that prior to failure, the contact network is essentially static, with only minor rearrangements of the particles.  However, the force network, which includes not only the contact information, but also the information 
about the interaction strength, undergoes a nontrivial evolution. 
An initial insight is reached by considering the relative
magnitudes of normal and tangential forces between the particles, and in particular the proportion of contacts that reach
Coulomb threshold.  More detailed understanding of the processes leading to failure is reached by the analysis of 
both spatial and temporal properties of the force network using the tools of persistent homology. We find that the forces between the particles undergo intermittent temporal variations ahead of the failure. In addition to this temporal intermittency, the response of the force 
network is found to be spatially dependent and influenced by proximity to the intruder.  Furthermore, the response is modified
significantly by the interaction strength, with the relevant measures describing the response showing differing behavior for
the contacts characterized by large interaction forces.
 \end{abstract}

\maketitle

\section{Introduction}

The processes related to failure in soft solids have been explored for a long time, and in particular
during last couple of decades a significant progress has been achieved; excellent review articles
are available, discussing soft solids in general~\cite{falk_langer_2011}, failure in the context 
of fracture and avalanching~\cite{daub_carlson_2010}, and related issue in the context of glass transition of amorphous materials~\cite{berthier_2011}, 
among other research directions.  In the context of granular matter, the concept of failure is closely related to jamming, another topic that has 
produced a significant interest, see~\cite{liunagel01,hecke_2010,liu_nagel_2010,behringer_2018} for reviews. In particular, a close connection 
between particle scale information, including local contact networks, and macroscale rheology of sheared or compressed granular materials, 
has been established.  In addition, a point was made years ago~\cite{cates98,cates_1999} that not only contact networks per s\'e are important 
in governing the system response, but also their strength; this point will be of particular relevance in present work. 

In this paper, we focus on a system in which the connection between contact and force networks, and 
material failure, can be analyzed in detail. We consider an intruder, large compared to particles size,
that is initially buried in a granular column. The intruder is exposed to an upward pullout force which 
increases with time until the material fails, similarly as considered recently in two dimensions (2D) in 
experiments with photoelastic particles~\cite{zhang_epj17}, and even more recently in 
3D experiments~\cite{Jalali2020}. 
 To some degree, this pull-out setup may be considered as a reverse of the impact 
of an object onto the surface of granular material~\cite{durian_nat07,clark_prl12}, although one may wonder to 
which degree slowing down a moving intruder is similar to the failure due to an applied force on an initially static object.

After discussing general, system-wide response to applied pull-out force, we
will focus in particular on the question of particle-scale response, both from static and dynamic points of view. As pointed out above, even if the intruder and the granular particles are
stationary, there could still be changes in the network describing mutual interactions between the 
granular particles, particles and the intruder, as well as particles and the walls. 
In the present context, such a network is weighted by the strength of the force between the 
particles.  This strength can be quantified by either normal, tangential, or some 
combination of 
these force components.  By now it has been established that the properties of such force 
network are crucial for the purpose of developing better understanding of granular systems
in the context of analysis of static packs~\cite{zhang_softmatter_2014}, shear~\cite{azema2014,dijksman_2018}, compression~\cite{epl12,pre14,daniels_softmatter_2019}, wave propagation~\cite{daniels_pre12,bassett_softmatter_2015}, or impact~\cite{pre18_impact}; see also~\cite{behringer_2018} for a recent review.  
Reaching this understanding is however complicated by the complexity of the force networks 
due to a large amount of time-dependent data.  Obviously, some type of simplification is needed; in addition, 
one would like to use measures that are precisely defined and that allow for a meaningful comparison 
of the force networks at different time instances, since we will be also interested in a dynamic setting.

In recent years, significant progress has been made towards developing better understanding of force networks using 
a variety of tools, including force network ensemble analysis~\cite{snoeijer_prl04,tighe_sm10,sarkar_prl13}, 
statistics-based methods~\cite{peters05,tordesillas_pre10,tordesillas_bob_pre12,makse_softmatter_2014}, 
network type of analysis~\cite{daniels_pre12, herrera_pre11,walker_pre12,tordesillas_pre_15,giusti_pre16}, 
and application of topological methods, more precisely persistent homology~\cite{arevalo_pre10,arevalo_pre13,ardanza_pre14,epl12,pre13,physicaD14,pre14,Pugnaloni_2016,kondic_2016}; see~\cite{Papadopoulos_18,behringer_2018} for recent reviews. While different methods provide a complementary insight, we will focus here on persistent
homology based approach since it allows for significant data reduction, for formulation of objective 
measures describing the force network, as well as for precise comparison of different networks in a
dynamic setting. 
Such an approach was used with success to discuss force networks for the systems that were 
compressed~\cite{epl12,pre14}, vibrated~\cite{Pugnaloni_2016,kondic_2016}, or sheared~\cite{gameiro_prf_2020}, 
as well as for the analysis 
of experimental data~\cite{dijksman_2018,pre18_impact}.  For the problem considered in the present
paper, one crucial aspect of the method is that it allows for consideration of the strength of 
interaction between the particles. This will be an important quantity in the analysis of the results. 
The robustness (and simplicity) of the measures describing the considered network will allow 
us to obtain both more precise and more intuitive understanding of the connection between the system-wide failure 
and force networks properties.   

This manuscript is organized as follows.  We continue in Sec.~\ref{sec:methodology} 
by discussing the methods used, providing first the basic description of the system considered
in Sec.~\ref{sec:study}, and then describing briefly the discrete element techniques employed in 
simulations in Sec.~\ref{sec:DEM}, normalization and scaling in Sec~\ref{sec:scaling}, 
followed by again brief description of the persistence homology
techniques that allow for the analysis of time-dependent weighted networks in Sec.~\ref{sec:PH}.  
Then, in Sec.~\ref{sec:results},
we discuss the main results, starting in Sec.~\ref{sec:macro} with the  macro-scale results describing 
intruder dynamics, pullout force, as well as the force on the system boundaries.  In Sec.~\ref{sec:meso},
we analyze the force network, describing in detail the process of failure.  
Section~\ref{sec:conclusions} is devoted to the conclusions of the present
study and to the discussion of possible future research directions. 

\section{Methodology} \label{sec:methodology}
\subsection{General setup} 
\label{sec:study}
\begin{figure}[h]
    \centering
    \includegraphics[width=0.3\textwidth]{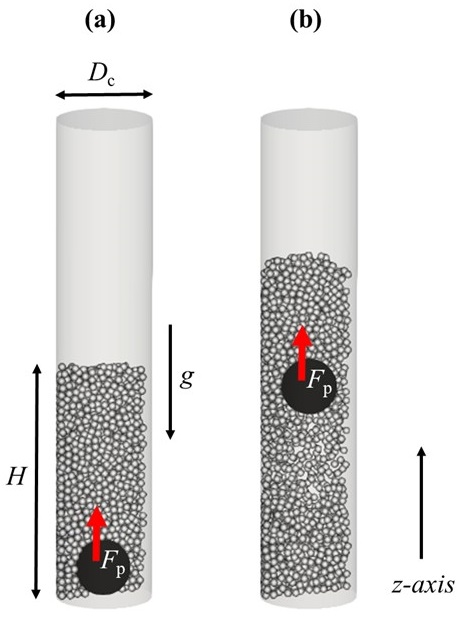} %
    \caption{(a) Initial configuration of particles filling the column with the intruder at the bottom. (b) A snapshot showing the configuration of particles and the intruder as it travels upwards after failure. Note that the particles in the front half of the column are not shown for a clear visualization. An animation of the pullout process showing the forces on particles is available as Supplementary Material~\cite{supp_mat_DEM}. }
    \label{fgr:Fig1}
\end{figure}

The simulation geometry consists of a vertical column of diameter $D_c$ filled with monodisperse spherical particles and a spherical intruder. The intruder is initially at contact with the center of the bottom of the column. Two column sizes are used, motivated by the 
ongoing experimental efforts~\cite{Jalali2020}.  The columns are filled with glass beads to a specified height, $H$, by pouring them randomly from the top at a certain mass flow rate. The initial configuration of particles and intruder is shown in Fig.~\ref{fgr:Fig1}.

After filling to a desired height, the granular material is relaxed until particles velocities are near zero.  
Once the particles come to a rest, the intruder is pulled vertically upward by a force, $F_\mathrm{p}$, which
is specified as a linear function of time with the slope of $1$ N/s. At a certain force, namely the pullout force at failure $F_\mathrm{pf}(H,D_\mathrm{c})$, the granular material fails and the intruder leaves the granular bed in fraction of a second.  
Figure~\ref{fgr:Fig1}b) shows a snapshot of the assembly where the intruder is halfway out of the granular bed; see also 
Supplementary Material~\cite{supp_mat_DEM}, showing an animation of the pull out process.  

\subsection{Computational approach}\label{sec:DEM}
\subsubsection{DEM solver}
We use the open source software, LIGGGHTS, for discrete element method (DEM) simulations of our granular system~\cite{Kloss2012}.  The DEM solves the equations of motion for individual particles and identifies the contacts and contact forces with neighboring particles. Applying the Newton’s second law for a particle $j$ with mass $m_\mathrm{j}$, the moment of inertia $I_\mathrm{j}$, translation velocity, $\vec{v}_\mathrm{j}$, and angular 
velocity, $\vec{\omega}_\mathrm{j}$, the equations of motion can be written as follows

\begin{equation}
    m_\mathrm{j} \frac{d \vec{v}_{\mathrm{j}}}{dt} = \sum_{k} \left(\vec{F}_\mathrm{n,jk} + \vec{F}_\mathrm{t,jk}  \right) + m_\mathrm{j} \vec{g},
    \nonumber
\end{equation}

\begin{equation}
    I_\mathrm{j} \frac{d \vec{\omega}_\mathrm{j}}{dt} = \sum_{k} \vec{T}_\mathrm{jk},
    \nonumber
\end{equation}
where $\vec{F}_\mathrm{n,jk}$ and $\vec{F}_\mathrm{t,jk}$ are the normal and tangential components of the contact forces between particles $j$ and $k$. $\vec{T}_\mathrm{jk}$ is the torque imposed by the tangential component of the contact force from particle $j$ on particle $k$, and $\vec{g}$ is the acceleration due to gravity. The normal and tangential forces contributing in building the two terms of stiffness and damping coefficients can be computed from the modified Hertz-Mindlin model as below\cite{Tsuji1992,Renzo2004}

\begin{equation}
   \vec{F}_\mathrm{n,jk} =  -\frac{4}{3}Y^{*}\sqrt{R^{*}}\delta_\mathrm{n}^\frac{3}{2} \vec{n}_\mathrm{jk} + 2 \sqrt{\frac{5}{6}}\beta \sqrt{S_\mathrm{n}m^{*}} \vec{v}_\mathrm{n,jk},
   \label{eq:fn}
\end{equation}

\begin{equation}
   \vec{F}_\mathrm{t,jk} =  -8G^{*}\sqrt{R^{*}\delta_\mathrm{n}} \vec{t}_\mathrm{jk} + 2 \sqrt{\frac{5}{6}}\beta \sqrt{S_\mathrm{t}m^{*}} \vec{v}_\mathrm{t,jk},
   \label{eq:ft}
\end{equation}
where $S_\mathrm{n}=2Y^{*}\sqrt{R^{*}\delta_\mathrm{n}}$ and $S_\mathrm{t}=8G^{*}\sqrt{R^{*}\delta_\mathrm{n}}$. Here, $\delta_\mathrm{n}$ is the normal overlap distance between the particles, $\vec{n}_\mathrm{jk}$ is the normal unit vector directed from the center of particle $j$ to that of particle $k$, $\vec{t}_\mathrm{jk}$ is the tangential displacement vector calculated by integrating the relative tangential velocity at the contact over time, $\vec{v}_\mathrm{n,jk}$ and $\vec{v}_\mathrm{t,jk}$ are the normal and tangential components of the relative velocity of particles $j$ and $k$. The expression for the torque is given as $\vec{T}_\mathrm{jk}=R_\mathrm{j}\vec{n}_\mathrm{jk} \times \vec{F}_\mathrm{t,jk}$. The magnitude of the tangential force is truncated by the Coulomb friction criterion: $\vec{F}_\mathrm{t,jk} \leq \mu\vec{F}_\mathrm{n,jk}$, where $\mu$ is the friction coefficient. In the 
above Eqs.~(\ref{eq:fn}~-~\ref{eq:ft}), the coefficient $\beta$,  expressed as a function of restitution coefficient $e$, is given by 
$    \beta = {\ln{e}/\sqrt{\ln^2\left(e\right)+\pi^2}}$.  
The other quantities used in Eqs.~(\ref{eq:fn}~-~\ref{eq:ft}) are given as follows

\begin{equation}
   \frac{1}{Y^{*}} = \frac{1-\nu_\mathrm{j}^2}{Y_\mathrm{j}} + \frac{1-\nu_\mathrm{k}^2}{Y_\mathrm{k}},
   \nonumber
\end{equation}

\begin{equation}
   \frac{1}{R^{*}} = \frac{1}{R_\mathrm{j}} + \frac{1}{R_\mathrm{k}}, \quad 
   \frac{1}{m^{*}} = \frac{1}{m_\mathrm{j}} + \frac{1}{m_\mathrm{k}}, 
   \nonumber
\end{equation}

\begin{equation}
   \frac{1}{G^{*}} = \frac{2\left(2-\nu_\mathrm{j}\right)\left(1+\nu_\mathrm{j}\right)}{Y_\mathrm{j}} + \frac{2\left(2-\nu_\mathrm{k}\right)\left(1+\nu_\mathrm{k}\right)}{Y_\mathrm{k}},
   \nonumber
\end{equation}
where $Y^{*}$ is the effective Young’s modulus, $G^{*}$ is the effective shear modulus, $\nu$ is the Poisson’s ratio, $R^{*}$ is the effective radius and $m^{*}$ is the effective mass. Similar equations as the above-mentioned ones are used to compute particle-wall contact forces, in which particle $k$ is assumed to be a particle with infinite radius and mass. 

\subsubsection{Simulation details}

\begin{table}[h]
\small
  \caption{\ Mechanical properties of materials and simulation parameters}
  \label{tbl:Table1}
  \begin{tabular*}{0.48\textwidth}{@{\extracolsep{\fill}}llll}
    \hline
    Mechanical & \hspace{-0.2 cm} Value & Mechanical &  \hspace{-0.3 cm} Value  \\
    properties &  & properties  &  \\
    \hline
    Young's modulus, GPa &  & Friction coefficient & \\
    \hspace {0.1 cm} intruder, bottom wall & \hspace{-0.2 cm} $200$ & \hspace {0.1 cm} intruder-glass beads, $\mu_{\rm ig}$ & \hspace{-0.3 cm} $0.45$\\
    \hspace {0.1 cm} glass beads & \hspace{-0.2 cm} $10$ & \hspace {0.1 cm} glass beads-glass beads, $\mu_{\rm gg}$ & \hspace{-0.3 cm} $0.4$\\
    \hspace {0.1 cm} sidewall & \hspace{-0.2 cm} $3$  & \hspace {0.1 cm} glass beads-sidewall, $\mu_{\rm gw}$ & \hspace{-0.3 cm} $0.3$\\\\
    Poisson's ratio &  & Restitution coefficient & \\
    \hspace {0.1 cm} intruder, bottom wall & \hspace{-0.2 cm} $0.3$ & \hspace {0.1 cm} intruder-glass beads & \hspace{-0.3 cm} $0.8$ \\
    \hspace {0.1 cm} glass beads & \hspace{-0.2 cm} $0.2$ & \hspace {0.1 cm} glass beads-glass beads & \hspace{-0.3 cm} $0.9$ \\
    \hspace {0.1 cm} sidewall & \hspace{-0.2 cm} $0.4$ & \hspace {0.1 cm} glass beads-sidewall & \hspace{-0.3 cm} $0.85$ \\\\
    \hline
    Simulation & \hspace{-0.2 cm} Value & Simulation &  \hspace{-0.3 cm} Value  \\
    parameters &  & parameters  &  \\
    \hline
    Density, kg/m$^3$ &  & Diameter, m &  \\
    \hspace {0.1 cm} intruder &  \hspace{-0.2 cm} $7728$ & \hspace{0.1 cm} intruder &  \hspace{-0.3 cm} $0.0349$\\
    \hspace {0.1 cm} glass beads & \hspace{-0.2 cm} $2500$ & \hspace{0.1 cm} glass beads  &  \hspace{-0.3 cm} $0.005$ \\
    Mass of intruder, kg & \hspace{-0.2 cm} $0.172$ & DEM time step, s  & \hspace{-0.3 cm} $5\times10^{-7}$ \\
    \hline
  \end{tabular*}
\end{table}

Table~\ref{tbl:Table1} specifies the mechanical properties of materials and simulation parameters. The mechanical properties of materials as given in Table~\ref{tbl:Table1} represent glass beads as the granular material, the sidewall of the column has properties close to PVC, and the intruder and bottom wall of the column have properties close to steel, motivated by recent experiments~\cite{Jalali2020}\footnote{Since the experiments, that will be reported elsewhere, were carried out using sand and involve huge number of particles, we do not discuss direct comparison in the present paper.}.  
The Young's modulus and the Poisson's ratio were obtained from literature~\cite{Tanaka2002,hostler_pre05_1,Koivisto2017}.  

The value of the restitution coefficient between glass beads is taken as $0.9$~\cite{Koivisto2017}. The restitution coefficient of steel is typically in the range of about $0.6-0.9$, and a value of $0.8$ is assigned here for steel and glass bead contacts. The restitution coefficient between glass beads and sidewall is assigned as $0.85$. The relevance of all above-mentioned values of the coefficient of restitution is found to be insignificant in present simulations as the granular material remains essentially in static state prior to failure. 

\subsection{Normalization of relevant quantities}\label{sec:scaling}

To obtain a better insight into the physics of the studied system, and for simplicity, we normalize the 
relevant quantities using the intruder diameter, $D_{\mathrm{i}}$ and the reference force, 
${F_\mathrm{0}}$.  The choice of 
$F_{\rm 0}$ is based on the weight of a reference volume, which is taken as a cylinder 
of diameter $D_{\rm i}$ and height $D_{\rm i}$, filled with granular matter.  
Since the volume fraction changes (very slightly) with filling height, $H$, due to gravitational compaction,
we choose the average value obtained in simulations (that use different $H$), giving the value of $F_{\rm 0} = 0.425$ N. 
We will see later in Sec.~\ref{sec:macro} that the force scale chosen as specified is natural for the considered problem.   
From this point, all the results are given using $D_{\rm i}$ and $F_{\rm 0}$ as the scales for the length 
and force, except if specified differently.   At a couple of places we will also need a time scale, which 
for simplicity we take as $\sqrt{D_{\rm i}/g}$.   

We note that the pullout force, $F_{\rm p}$, does not include the weight of intruder. Therefore, the 
value of $F_{\rm p} = 0$ corresponds to the time at which the applied force on the intruder is
exactly equal to its weight.  

\subsection{Persistent homology}\label{sec:PH}

For the present purposes, persistent homology could be thought of a tool for describing a complicated weighted network 
(such as the one describing interaction forces between the particles) in a form of diagrams, so called persistent 
diagrams ($\pd$s). These diagrams are obtained by filtration, or thresholding, the strength of the interactions between the particles.  
As an example, let us assume that this strength is quantified by the normal interaction force between the particles in the 
system.  Then, the simplest persistence diagrams, which we refer to as $\beta_0$ $\pd$ for the reasons that will become 
clear shortly, essentially traces how `structures' (could be thought of as `force chains', without attempting to define 
them) appear as a filtration level is decreased, or disappear as two structures merge. Such a $\pd$ contains rather 
complete information about how connectivity between the contacts depends on their strength.  Figure~\ref{fig:PD} shows examples 
of the $\pd$s computed at a given time for the system considered in the present paper (both for $\beta_0$ discussed above, 
and for $\beta_1$, that describes loops or cycles, and discussed further below).  Each point (generator) in this 
diagram has two coordinates, `birth' - specifying when (at which filtration level) it appeared, and `death' - specifying 
when it disappeared, due to merging with another structure.  Note that the term `persistence' in this context refers to the 
force range over which a certain structure `persists', or survives, as a considered filtration level is changed.   
The interested reader could find rather detailed description on 
how these diagrams can be (properly) defined and computed in~\cite{physicaD14}, and a simplified description including
toy examples and insightful animations in~\cite{pre14}.   For the purposes of the present paper, we list here
few key features of $\pd$s that are important for understanding their meaning, relevance, and use:
\begin{figure}[h]
\centering
  \includegraphics[width=0.475\textwidth]{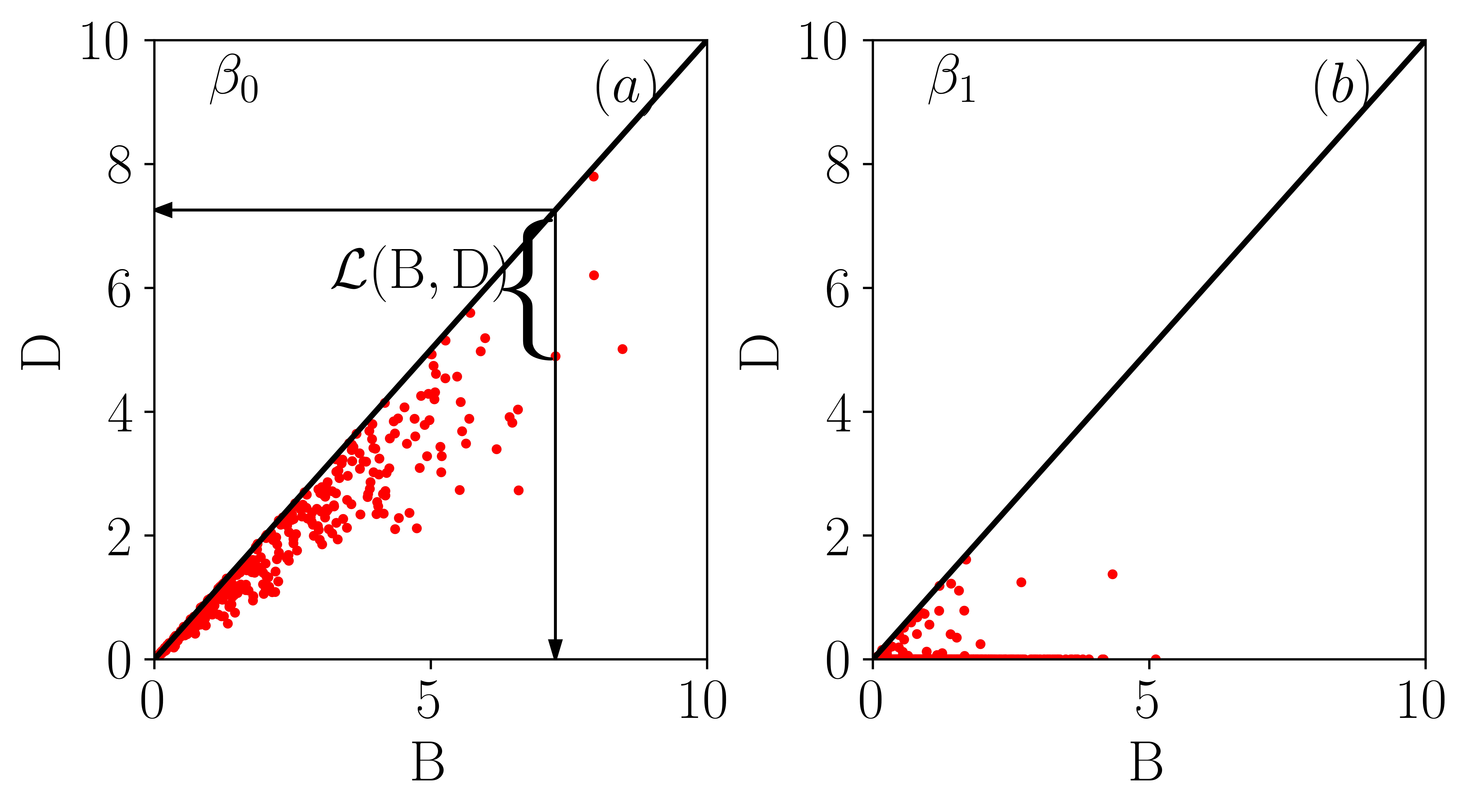}
  \caption{Examples of persistence diagrams, $\pd$s,  for a) components (clusters), $\beta_0$, and  b) for loops
  (cycles), $\beta_1$.  The axes show the force level at which a structure appears (B) and disappears (D)  
  The lifespan ${\cal L} =$ B-D  illustrates the force threshold range over which a a structure persists.  
    These  particular $\pd$s were obtained for the reference case discussed later in the text (circled data point in Fig.~\ref{fgr:Fig3}), and for pulling force 
  $F_{\rm p} = 9$ (dimensionless value, as discussed in Sec.~\ref{sec:scaling}).  
  See Supplementary Material for the animations showing full evolution of the $\pd$s as $F_{\rm p}$
  increases and failure is approached~\cite{supp_mat_PD}.
  }
  \label{fig:PD}
\end{figure}
\begin{itemize}
    \item Persistent diagrams live in a metric space and therefore can be compared in a meaningful manner.  This means 
    that one can compare the weighted force network at different points in time, which is a crucial feature if one 
    wants to understand an evolving system. 
    \item Since the description of a weighted force network in terms of persistence diagrams provides a significant data reduction, there is an associated loss of information, and therefore $\pd$s do not provide full information about the underlying
    network.  However, the information contained in $\pd$s  still provides for all practical purposes 
    a reasonably precise insight not only regarding the number of
    structures, but about how their connectivity changes as one considers different thresholding (filtering) levels.  
    \item The number of $\pd$s for a given weighted network is equal to the number of spatial dimensions:
    i.e., for a 3D system considered here, in principle there are three $\pd$s that can be
    associated: $\beta_0$ $\pd$ describing clusters; $\beta_1$ $\pd$ describing loops (cycles), 
    and $\beta_2$ $\pd$ describing three dimensional holes.  In the present work, we do 
    not find $\beta_2$'s, and therefore focus on 
    $\beta_0$ and $\beta_1$ $\pd$s.  Figure~\ref{fig:PD} shows that the $\beta_0$ generators 
    appear on higher force level than $\beta_1$ ones, since $\beta_0$ structures need to merge to form loops, and this happens on lower force levels.   The points at the death level ${\rm D} =0$ in Fig.~\ref{fig:PD}b) show all the loops that formed, but never died: a loop dies at the threshold level D when it is filled up with contacts with the force of at least D, and for many loops this does not happen; see~\cite{physicaD14} for more details.  On the other hand, all $\beta_0$ components disappear at nonzero D, since for low values of force all separate clusters merge.
    \item A $\pd$ provides information about a weighted network that is threshold independent: there is no need to specify
    a threshold level to describe the considered network.  A $\pd$ provides information about all threshold levels at once.  This is a major difference compared to other measures describing force networks, which often require separation 
    of a force network into a `strong' or a `weak' network.  With this being said, one can in principle 
    consider only a subset of points in a $\pd$, and focus on a particular range of interest.   For example, for the 
    diagram shown in Fig.~\ref{fig:PD}, one could consider only the points found for large thresholds, 
    describing a strong part of the force network, or the points for small thresholds, describing a weak part of the 
    network.    
    \item The number of `force chains' or `clusters' of connected contacts (or the number of loops (cycles)) 
    characterized by a force larger than a specified threshold can be easily extracted from the $\pd$s, 
    providing therefore information about Betti numbers, $\beta_n$, that essentially measure a number of features of a given type. 
    Note however that the number itself does not provide any information about connectivity.  Furthermore, 
    one can show~\cite{physicaD14} that Betti numbers are susceptible to noise, meaning that a small change in the input
    data could lead to a large change of the Betti numbers.  
    Finally, the $\beta_n$'s are threshold dependent, and therefore provide less
    complete information about the underlying weighted network.  
    \item The $\pd$s are essentially point clouds, and an appropriate measure for their quantification needs to be developed. One
    possibility is based on the idea that both the number of points in a diagram, and for how long (that is, for 
    how many thresholds levels) a point persists, are appropriate measures describing the 
    force network between particles.  Using landscape as an analogy, the number of points in $\beta_0$ $\pd$ specifies the number of (mountain) peaks, and their lifespan (see below) describes how well developed these peaks are, or, to push the landscape analogy further, how high they are compared to the `valleys' that surround them.  The concept of lifespan, $\mathcal{L}$ is illustrated in Fig.~\ref{fig:PD}, as essentially the difference between birth (B) and 
    death, D, coordinates, so $\mathcal{L} = {\rm B}  -{\rm D}$.  Both measures could be combined into one by defining total 
    persistence, TP, as a sum of all lifespans.   We will be using TP in discussing some properties of the force      
    networks in the considered system. 
    \item The reader should note the enormity of the data reduction described so far: 
    first a weighted force network is reduced to a point cloud ($\pd$) and then this point cloud to a single 
    number (TP).  Such a reduction clearly leads to huge data loss, but as we will see, still provides an insightful 
    information about the underlying weighted network.   
\end{itemize}

$\pd$s provide information about the state of the system at a given time, and do not include any information about the dynamics. 
As pointed out, however, $\pd$s can be compared, and in particular the concept of {\it distance} between $\pd$s is 
well defined~\cite{physicaD14}.   The distance between two $\pd$s could be thought as a minimum (Euclidian) distance
by which the points in one diagram need to be adjusted to map them exactly to the other one; if the number of points 
in two $\pd$s is different, then extra points are mapped to the diagonal. 
Stated formally
\begin{equation}
d_{W^q}(\pd,\pd') = \inf_{\gamma\colon \pd \to \pd'} \left(\sum_{p\in\pd} \| p -\gamma(p)\|^q_\infty \right)^{1/q}. 
\label{eq:dist}
\end{equation}
Here, $d_{W^q}$ stands for the degree q Wasserstein distance, and $\pd$, $\pd'$ are the two considered $\pd$s. Computing 
the distance is a computationally expensive process in particular for complex diagrams with many points.  In the 
present work, we focus on $q=2$ for simplicity, and perform calculations using the method discussed in Refs.~\cite{gudhi,rTDA}.

\section{Results and discussion}\label{sec:results}

We discuss main findings in this section, focusing first in Sec.~\ref{sec:macro} on the system-scale
response, including the force at the failure point, the influence of system boundaries, 
geometry, and material parameters.  
These results set a stage for Sec.~\ref{sec:meso}, where more in-depth analysis of the failure 
process itself is discussed by considering the weighted force networks that form spontaneously 
between granular particles.


\subsection{Failure: Macro-scale picture}\label{sec:macro}

\subsubsection{Failure force} 

\begin{figure}[h]
\centering
  \includegraphics[width=0.475\textwidth]{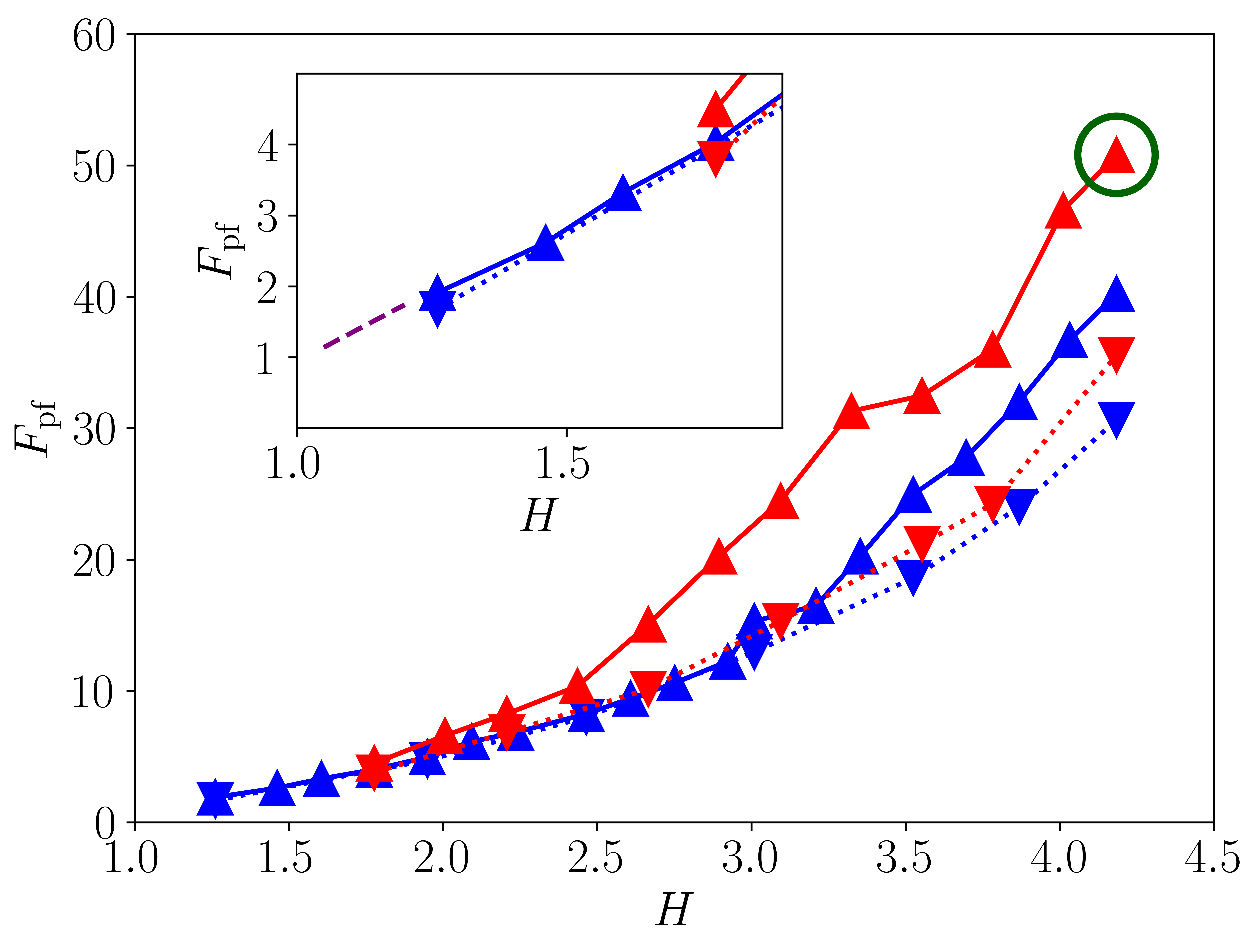}
  \caption{The pullout force at failure, $F_{\rm pf}$, for different filling heights, $H$, 
  and for two different column diameters, \textcolor{red}{(red), $D_{\rm c} = 1.49$}, and 
    \textcolor{blue}{(blue) $D_{\rm c} = 1.75$}.
  $\blacktriangle$ and $\blacktriangledown$ corresponding to the higher (given in Table~\ref{tbl:Table1}) and 
  lower (discussed in the text) values of the friction coefficients, 
  respectively.   The circled data point shows the reference case that is analyzed in detail later in the paper. 
  The inset expands the region for small $H$, showing that
  $F_{\rm pf} \rightarrow 1$ as $H\rightarrow 1$ (\textcolor{purple}{(purple) dashed line}).  
  Note that all quantities in this and the following figures are dimensionless. 
 }
  \label{fgr:Fig3}
\end{figure}

We start by discussing the magnitude of the pulling force applied to the intruder that causes
failure. Figure~\ref*{fgr:Fig3} shows how the pullout force at failure, $F_{\rm pf}$, depends on the filling 
height, $H$,  for the two granular columns; at first we focus on the $\blacktriangle$ symbols (connected by the solid lines) which show the results obtained using the friction parameters given 
in Table~\ref{tbl:Table1}.

We see that $F_{\rm pf}$ increases with $H$, as expected, since larger $H$ leads to a larger pressure on the intruder. 
Figure~\ref*{fgr:Fig3} (inset) also clarifies the force scaling: with the
choice of $F_0$ specified in Sec.~\ref{sec:scaling}, $F_{\rm pf}$ extrapolates to $F_{\rm pf} \approx 1$ as $H\rightarrow 1$ (note
that all quantities are dimensionless).  Furthermore, we observe that $F_{\rm pf}$ is larger for the smaller diameter column, showing the relevance of particle-wall interactions.  This finding can be understood based on an increased
forces on the side walls (Janssen effect, see~\cite{ciamarra_prl20} for a recent 
discussion), which are considered in more detail in Sec.~\ref{sec:parameters}. 
Note that the effect of sidewall on $F_{\rm pf}$ diminishes as $H$ decreases, 
since for small $H$ we enter the hydrostatic regime where the column size is not relevant.

One important question involves the influence of friction coefficient on the pullout failure force.  In the 
present DEM simulations, there are three relevant friction coefficients: intruder - glass beads, $\mu_{\rm ig}$, 
glass beads - sidewall, $\mu_{\rm gw}$, and glass beads - glass beads, $\mu_{\rm gg}$.  
Figure~\ref*{fgr:Fig3} shows the influence of friction: $\blacktriangle$ were obtained 
using the values of friction coefficients given in Table~\ref{tbl:Table1}, and the values shown by $\blacktriangle$ were obtained for lower values of friction coefficients 
($\mu_{\rm ig} = 0.4$, $ \mu_{\rm gg} = 0.35$, and $\mu_{\rm gw} = 0.25$).  
One could wonder which of the three friction coefficients is the most/least relevant; to answer this question, 
we discuss the reference case (circled data point in Fig.~\ref{fgr:Fig3}) in more detail. 
The $F_{\rm pf}$ for the reference case is $50.8$. When 
only $\mu_{\rm ig}$ is reduced, we find $F_{\rm pf} =  50.1$; 
when $\mu_{\rm gg}$ is reduced, $F_{\rm pf} = 46.6$, and when $\mu_{gw}$ is reduced, $F_{\rm pf} = 43.8$.   Thus, the effect caused by the change in particle-wall friction coefficient is the most dominant, although all friction 
coefficients influence $F_{\rm pf}$.  We note that these variations are considerably
larger than the ones found between different realizations of nominally the same simulation, when initial particle configurations were modified.  This variability is less than 1\%.  

The relevance of particle-wall friction discussed above also suggests a possible explanation for a slightly 
weaker influence of friction for a larger column diameter, $D_{\rm c}$.  Furthermore, the influence of friction disappears
for small values of $H$.  This finding shows that in the hydrostatic regime frictional effects do not 
influence the failure process.

\subsubsection{Parameter study}
\label{sec:parameters}

\begin{figure}[thb]
\centering
  \includegraphics[width=0.45\textwidth]{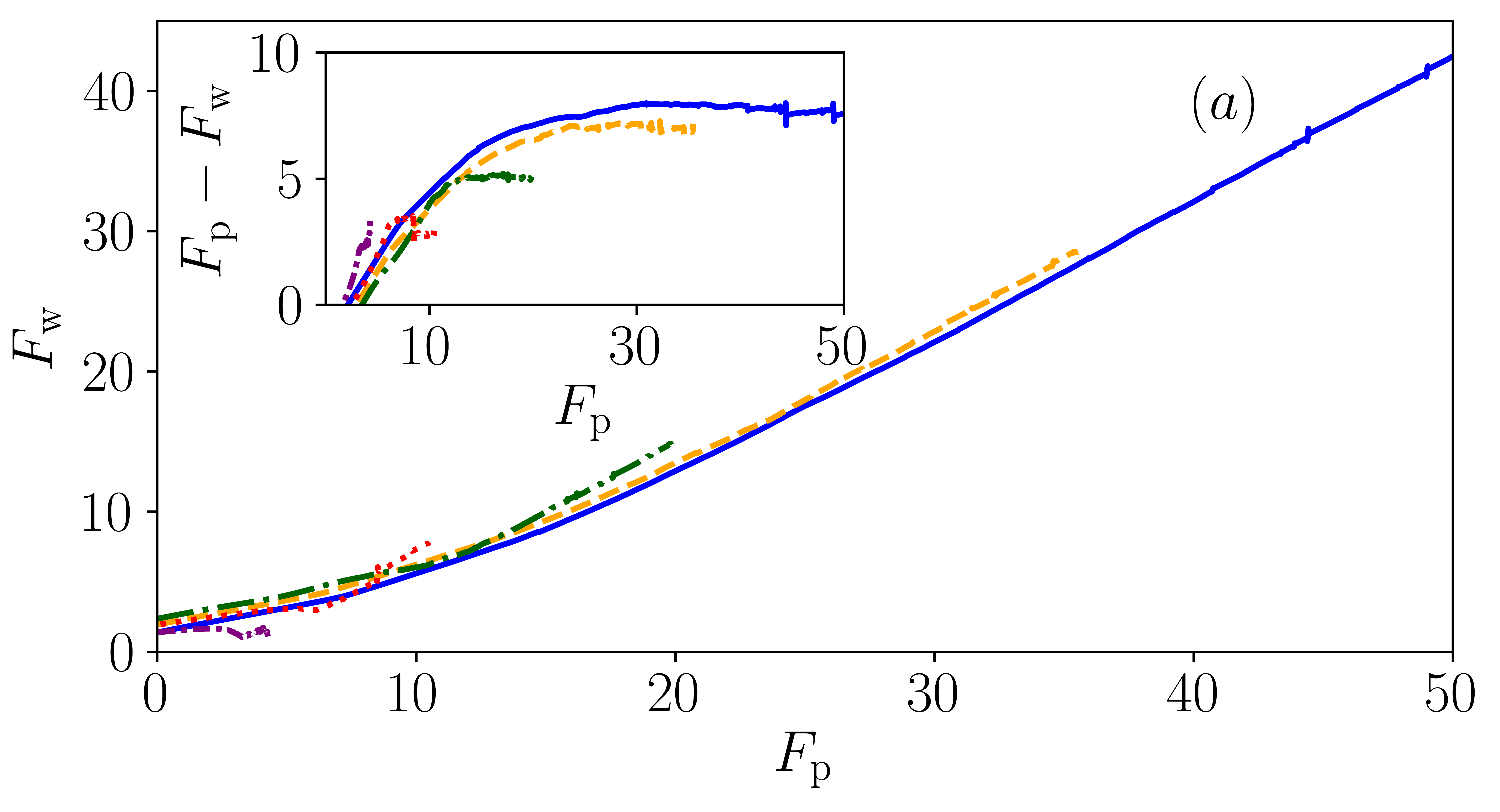}
    \includegraphics[width=0.45\textwidth]{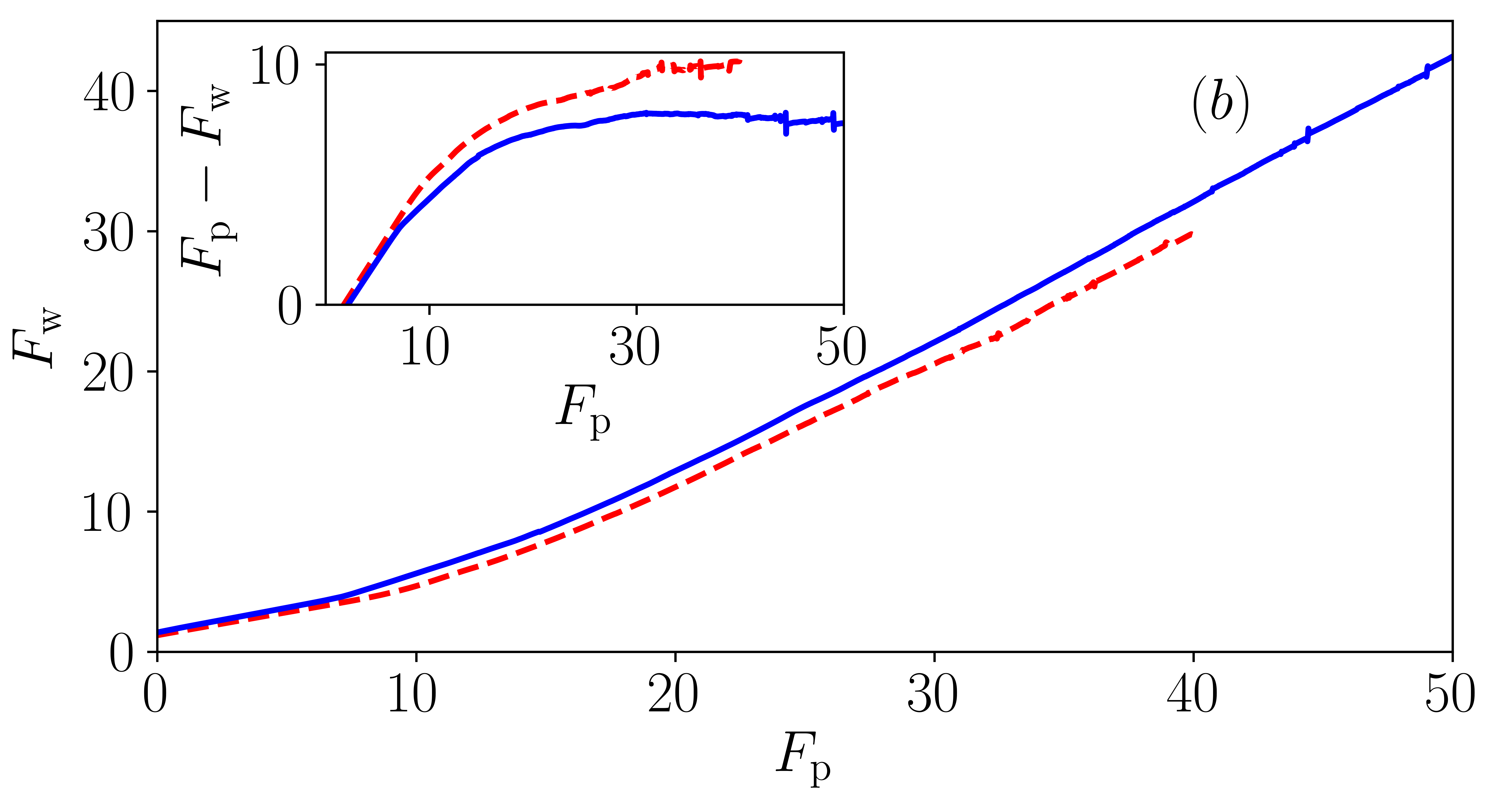}
   \caption{The sidewall force, $F_{\rm w}$ in the (vertical) $z$ direction exerted by the particles.
   (a) $D_c = 1.49$ with variation of the filling height: \textcolor{blue}{$H=4.18$ (solid blue line)}, \textcolor{orange}{$H=3.78$ (dashed orange line)}, \textcolor{green}{$H=2.89$ (dash-dot green line)}, \textcolor{red}{$H=2.21$ (dotted red line)}, and \textcolor{purple}{$H=1.78$ (loosely dashed purple
   line)}.
  (b) $H = 4.18$ with variation of the column diameter: \textcolor{blue}{$D_{\rm c}=1.49$ (solid blue line)} and 
  \textcolor{red}{$D_{\rm c}=1.75$ (dashed red line)}.
   }
  \label{fgr:Fw}
\end{figure}

Here we compare the results obtained using the smaller column diameter, and varying the 
height of the granular material, $H$. 
Figure~\ref{fgr:Fw}a) shows the (vertical) $z$-component of the force exerted by particles on the sidewall,  
$F_{\rm w}$,  versus the pullout force, $F_\mathrm{p}$, as $H$ is varied.   The reported values of $F_{\rm w}$
are obtained by averaging the instantaneous values over a short time window (of duration $t \approx 0.13$).

We focus at first on the results obtained for large filling heights in Fig.~\ref{fgr:Fw}a).  Clearly
$F_{\rm w}$ grows with $F_{\rm p}$, showing that the pullout force is transferred via particles to the sidewall. 
This growth is non-trivial however: as $F_{\rm p}$ increases, the difference $F_{\rm p} - F_{\rm w}$, shown in the inset of Fig.~\ref{fgr:Fw}a), 
reaches a plateau whose height depends on $H$.  Therefore, for small $F_{\rm p}$, the
coupling of $F_{\rm p}$ and $F_{\rm w}$ is not complete, with $F_{\rm p}$ increasing faster than 
$F_{\rm w}$.  However, for larger values of $F_{\rm p}$, the coupling becomes stronger, with 
essentially all the applied force transferred to the sidewall. The fact that the height of the plateau
is $H$-dependent suggests that the forces between the particles (that may depend on 
both $H$ and $F_{\rm p}$) play a role.   We defer further discussion of this effect until considering these forces in 
Sec.~\ref{sec:meso}.  

For small 
filling heights ($H = 1.78$ and $H = 2.21$ in Fig.~\ref{fgr:Fw}a)) the results
are less clear, with non-monotonous growth of $F_{\rm p}$ and $F_{\rm p} - F_{\rm w}$ as $H$ is increased; possibly in 
this regime inverse Janssen effect discussed recently~\cite{ciamarra_prl20} becomes relevant.   We leave detailed 
discussion of this regime for future work. 
 
Next, we focus on comparing the results of simulations carried out using 
different column diameters, $D_{\rm c}$, with the height, $H$, 
kept fixed. Figure~\ref{fgr:Fw}b) shows the results: 
we observe the $F_{\rm w}$ depends on $D_{\rm c}$, with 
the larger magnitude of $F_{\rm w}$ at 
the failure  for smaller $D_{\rm c}$.   The inset in Fig.~\ref{fgr:Fw}b), which
plots the difference $F_{\rm p} - F_{\rm w}$, shows that the height of the plateau at which this 
difference saturates increases as $D_c$ increases, and the plateau itself becomes less 
clearly defined.  This again shows that for larger column diameters, the wall forces 
become less important in helping the granular material to avoid failure.  
This motivates us to analyse further the response of the 
force networks prior to the failure.  This is discussed in the next section.

\subsection{Road to failure} 
 \label{sec:meso}
 
 In this section we discuss the failure process itself, focusing on a single 
 choice of parameters, represented by the reference case (circled data point in Fig.~\ref{fgr:Fig3}).
To analyze the failure, we use the measures that are appropriate for discussion of the static and 
dynamic properties of the system,  starting from the classical ones, such as forces, 
energies, and particle contacts, and then focusing on the 
 force networks between the particles and the particles and the intruder.

\begin{figure}[ht!]
\centering
  \includegraphics[width=0.475\textwidth]{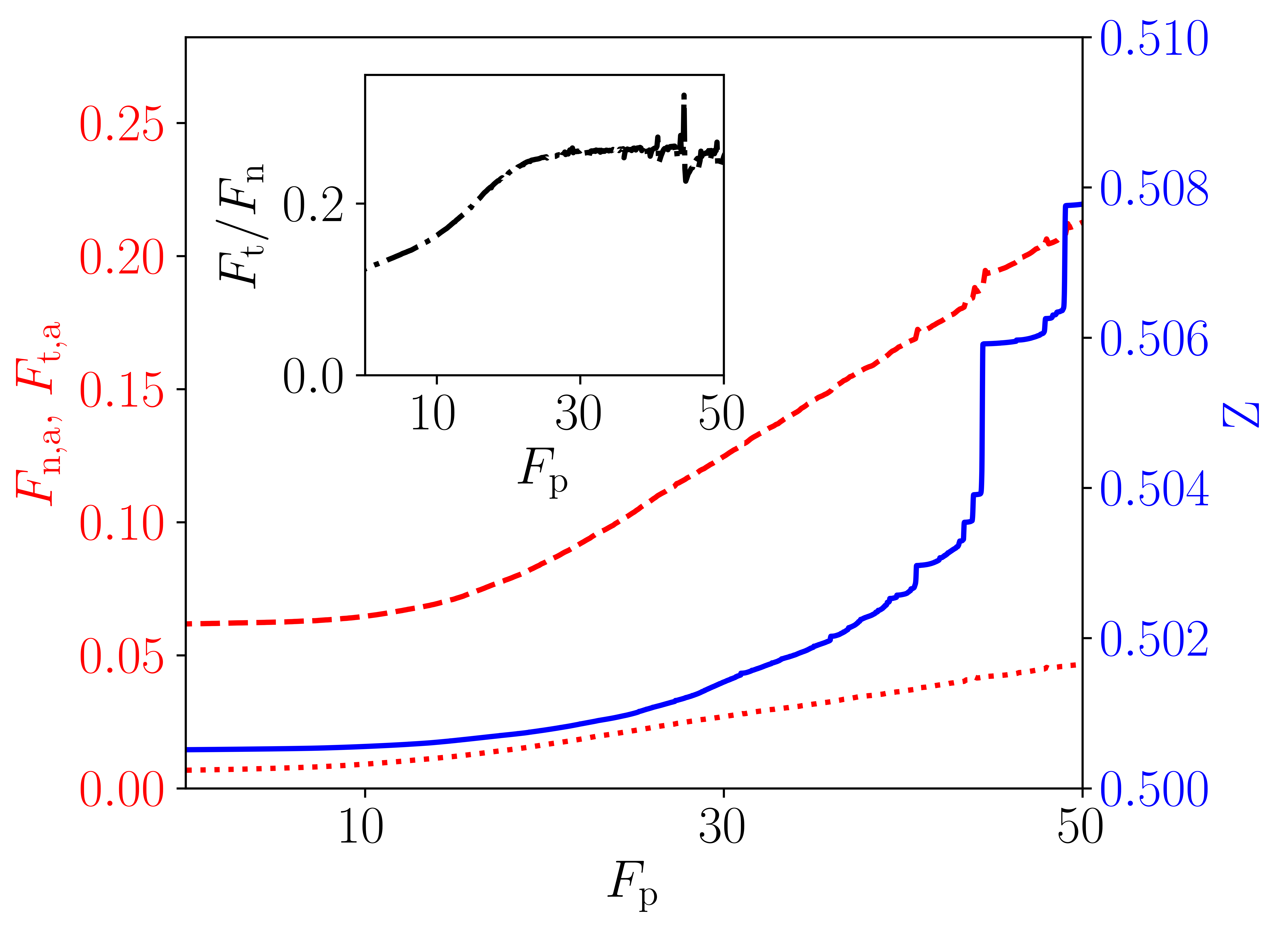}
    \caption{Average \textcolor{red}{normal (dashed red line)} and \textcolor{red}{tangential (dotted red line)} forces, 
    \textcolor{red}{$F_{\rm n,a},~F_{\rm t,a}$}, respectively, between the granular particles, 
     and the particles and the intruder \textcolor{red}{(left vertical axis)};  
     \textcolor{blue}{$z$-position (solid blue line)} of the intruder \textcolor{blue}{(right vertical axis)};  
     Inset: average ratio of $F_{\rm t}/F_{\rm n}$.
   }
  \label{fig:force}
\end{figure}

\subsubsection{Forces, Energies, Contacts}
\label{sec:global}

Figure~\ref{fig:force} shows the normal and tangential average forces, $F_{\rm n,a},~F_{\rm t,a}$,  respectively, 
between the particles (and the particles and the intruder), as well as the intruder's $z$-position.
We focus on the time period for which the pulling force is positive (after subtracting scaled intruder's weight). 
Figure~\ref{fig:force} shows that the dynamics of the intruder is minimal before the granular system fails: note
that the whole range on the vertical $z$ axis is just 1\% of the intruder's diameter (see also~\cite{supp_mat_PD}).
While for smaller values of $F_{\rm p}$, $F_{\rm n, a}$ does not change much, for 
larger $F_{\rm p}$'s, $F_{\rm n, a}$ grows approximately linearly with $F_{\rm p}$.  Since 
the change of $F_{\rm n, a}$ is considerable, in what follows we will use this time-dependent
value of $F_{\rm n, a}$ for normalization of the results discussing properties of the force network.  

Tangential force, $F_{\rm t,a}$, also shown in Fig.~\ref{fig:force}, provides relevant additional information.  While
$F_{\rm t,a}$ is smaller than $F_{\rm n,a}$ (being limited by the Coulomb threshold), the ratio
$F_{\rm t,a}/F_{\rm n,a}$ grows as failure is approached.  
This result provides an initial insight regarding failure:
the particle contacts get increasingly loaded by the tangential force, and approach sliding.  Significant
amount of sliding contacts appears as a precursor to failure.  
The sliding effect is also confirmed by the inset which shows the average 
of $F_{\rm t}/F_{\rm n}$ value at each contact (note that
this quantity is not necessarily the same as $F_{\rm t,a}/F_{\rm n,a}$).  We also comment on 
the similarities of the insets of Figs.~\ref{fig:force} and~\ref{fgr:Fw}, with both
showing similar qualitative behavior: this is not surprising since the tangential inter-particle
force clearly plays a role in determining $F_{\rm w}$.  For larger filling height, $H$, the normal forces between
the granular particles will be larger due to gravitational effects and therefore the
allowed range for tangential forces would be larger as well, leading to stronger wall force, 
consistently with the results shown in the inset of Fig.~\ref{fgr:Fw}(a).

\begin{figure}[ht]
\centering
  \includegraphics[width=0.475\textwidth]{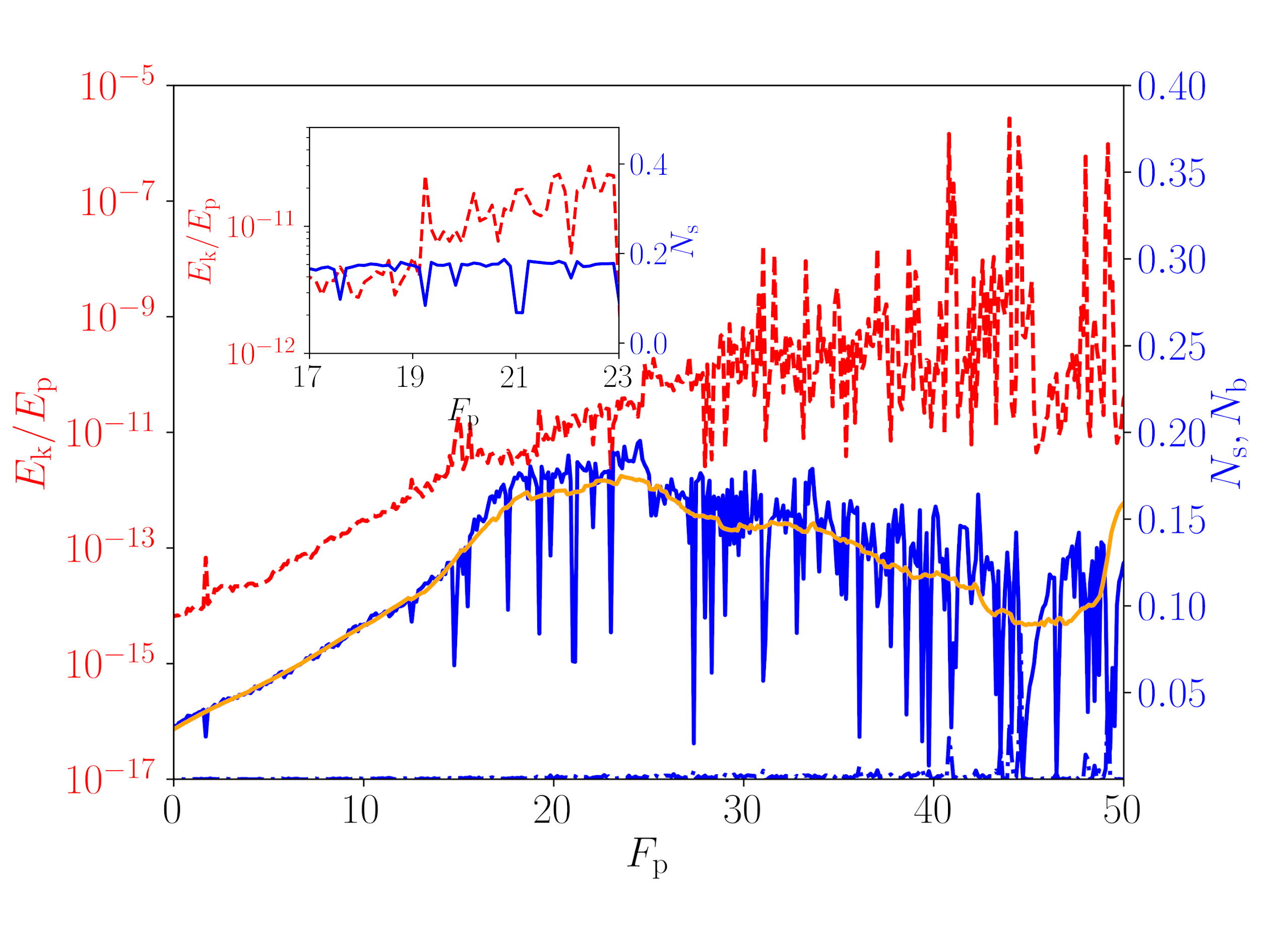}
     \caption{(left) \textcolor{red}{$E_{\rm k} / E_{\rm p} $ (dashed red line)}: the ratio of kinetic and potential energies of the  particles; 
     (right) \textcolor{blue}{$N_{\rm s}$ (solid blue line)}: the ratio of the number of contacts reaching Coulomb threshold and the total number
     of contacts, the \textcolor{orange}{orange solid line} shows the running average of \textcolor{blue}{$N_{\rm s}$};  
     \textcolor{blue}{$N_{\rm b}$ (dash-doted blue line)}: the ratio of the number of broken contacts and the total number of contacts.  
     Inset: expanded view showing \textcolor{red}{$E_{\rm k} / E_{\rm p} $} and \textcolor{blue}{$N_{\rm s}$} illustrating frequent
     anti-correlation.
   }
  \label{fig:energy}
\end{figure}

Continuing the discussion of sliding contacts, 
Fig.~\ref{fig:energy} plots three quantities of interest: $N_{\rm s}$, the ratio of the number of contacts that have 
reached Coulomb threshold (sliding contacts) and the total number of contacts;  $N_{\rm b}$, the ratio of the number 
of broken contacts and the total number of contacts, 
and the ratio of kinetic and potential energies of the granular 
particles, $E_{\rm k}/E_{\rm p}$ (here $E_{\rm k}$ includes only translation degrees of freedom, and 
$E_{\rm p}$ is measured relative to the bottom of the column).  All quantities are calculated
over the time intervals of duration $\Delta t \approx 0.84$. While the results are rather noisy, we can still reach some relevant conclusions, which could be also 
verified by inspecting the inset of Fig.~\ref{fig:energy}, which shows an expanded view of 
the energy ratio and $N_{\rm s}$. First of all, very small values of $E_{\rm k}/E_{\rm p}$ suggest 
minimal rearrangements of the particles during the period before failure, consistently with small $N_{\rm b}$
(note that the latter quantity is barely visible in Fig.~\ref{fig:energy}, showing that only small fraction 
of contacts breaks before failure).  While detailed inspection of the data does not show any 
correlation between $N_{\rm b}$ and $E_{\rm k}/E_{\rm p}$, the same conclusion does not apply to 
$N_{\rm s}$, for which we find significant (anti)correlation with $E_{\rm k}/E_{\rm p}$.  This result shows that, 
while reaching Coulomb threshold may not lead to breaking of a contact, it may lead to sliding, which reflects itself 
in an increase of kinetic energy. 

The results presented so far provide a basic idea about the processes leading to failure: the tangential force increases
at particle-particle and particle-intruder contacts faster than the normal force; the number of sliding contacts increases
as well, until at failure point the Coulomb threshold is overcome and the intruder starts to move.  Until failure 
occurs, the intruder is essentially static, with the change of position measured as a small fractions of the intruder diameter.  
Additional insight can be reached by considering spatial and temporal properties of the 
force network, as discussed next. 

\subsubsection{Force network and failure}
\label{sec:networks}

The tools of persistent homology, discussed briefly in Sec.~\ref{sec:PH}, provide an extensive information about the 
force networks.  We start by discussing rather straightforward measures focusing on the structures
(clusters or loops) introduced in Sec.~\ref{sec:PH}.    For brevity, we focus in this section on discussing 
the measures computed using the normal force, $F_{\rm n}$, only;  the findings obtained by considering the tangential
forces are found to provide consistent information with the one reported here. 

\begin{figure}[ht]
\centering
  \includegraphics[width=0.475\textwidth]{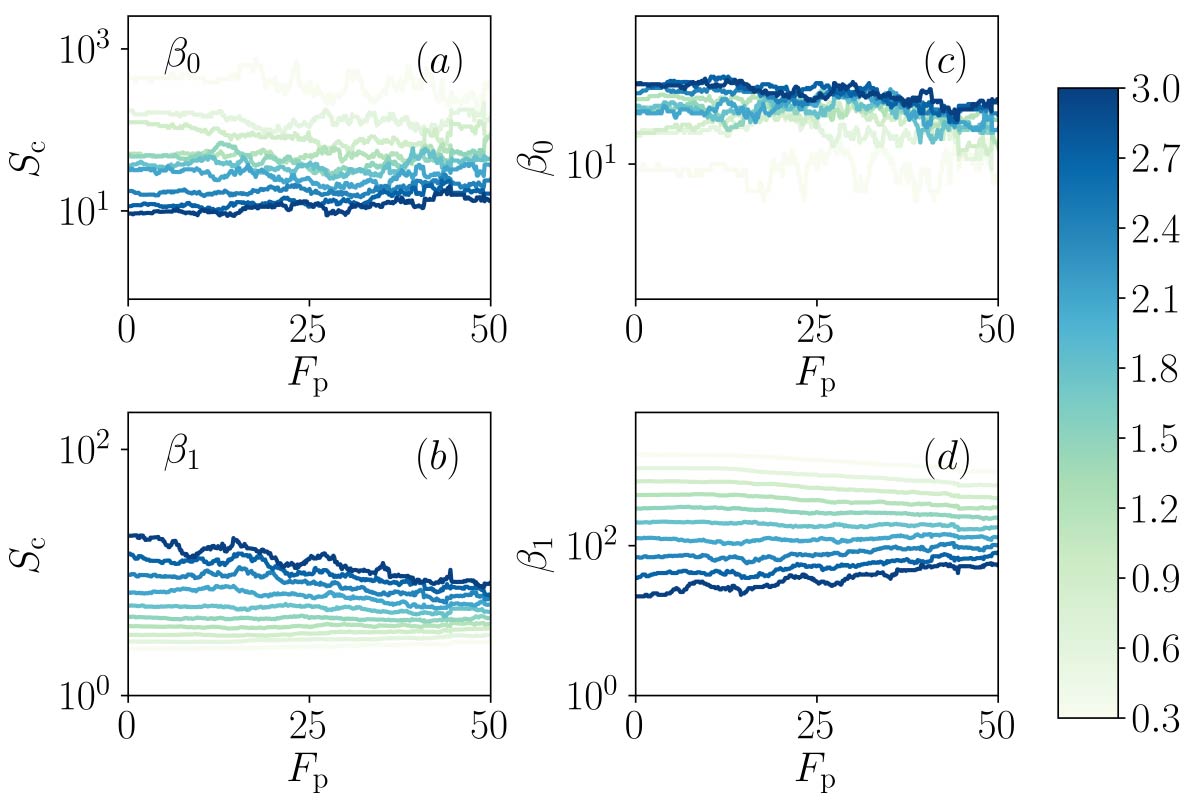}
    \caption{(a, c) $\beta_{\rm 0}$ component (`force chains') size (a) and number (c); 
   (b, d) $\beta_{\rm 1}$ component (loops, `cycles') size (b) and number (d) for different force threshold 
   values (shown by the color bar).  
   Note the opposing trend of $\beta_{\rm 0}$ and $\beta_{\rm 1}$; e.~g., Betti numbers
   for the clusters ($\beta_{\rm 0}$) increase for the considered thresholds since for small 
   threshold values all clusters are connected; the loops ($\beta_{\rm 1}$) (d) are more numerous
   for small threshold values due to merging of the clusters.  Note the use of log scale on the 
   vertical axes.  
   }
  \label{fig:clusters}
\end{figure}

Figure~\ref{fig:clusters} shows the results for the $\beta_{\rm 0}$ (a, c), and $\beta_{\rm 1}$ (b, d), 
both for the size, $S_{\rm c}$ (a, b) and the (Betti) numbers (c, d) of the structures.   These quantities can be easily
extracted from the corresponding persistence diagrams: Betti numbers, $\beta_{\rm 0},~\beta_{\rm 1}$ could be found by subtracting the number of components that die from the number of components that are born above specified threshold, and $S_{\rm c}$
is simply the number of contacts (with force above chosen threshold) divided by $\beta_{\rm 0}$ or $\beta_{\rm 1}$.  
Consistently with the results discussed in Sec.~\ref{sec:global}, 
Fig.~\ref{fig:clusters} shows that the force network evolves even while the 
intruder is almost stationary, and the contact network is essentially unperturbed.  This evolution is 
particularly obvious when considering $\beta_{\rm 1}$ structures for the large thresholds values: the (average) size 
of $\beta_{\rm 1}$ structures increases significantly as $F_{\rm p}$ increases, see Fig.~\ref{fig:clusters}(b), 
while the corresponding $\beta_{\rm 1}$ decreases, Fig.~\ref{fig:clusters}(d), albeit not so fast
(note the log scale on the vertical axes).  Basically, what happens here is that 
the loops become smaller and their number grows as the pullout force increases.  The conjecture is that the 
loops play an important role in stabilizing the system exposed to external forcing, consistently with 
the findings reached by considering impact experiments~\cite{pre18_impact}, where loops were found to play a
significant role in slowing down an intruder entering the domain filled with particles. 
The changes in the size and number 
of $\beta_{\rm 0}$ structures are much more gradual, but consistent, showing (again for the large thresholds) 
a slow increase in the size and slow decrease in the number - essentially, the clusters merge together and 
encompass more and more
contacts as the pullout force increases. Regarding the threshold dependence, we note that the number of 
$\beta_0$ structures is an increasing function of chosen threshold for the range chosen: 
here we are considering the range of thresholds at which we still have many contacts (viz. the $\pd$ diagrams
shown in Fig.~\ref{fig:PD} and the corresponding animations 
in Supplementary Materials~\cite{supp_mat_DEM}); for even larger
thresholds, $\beta_0$ would decrease similarly as observed in a recent study of suspensions~\cite{gameiro_prf_2020}.
We also note that the sizes, $S_{\rm c}$, of $\beta_0$ and $\beta_1$ structures are comparable at large
thresholds, as expected.  We remind the reader that the force thresholds used in 
this figure are normalized by the current value of the average (normal) contact force plotted 
in Fig.~\ref{fig:force}.

\begin{figure}[ht]
\centering
  \includegraphics[width=0.475\textwidth]{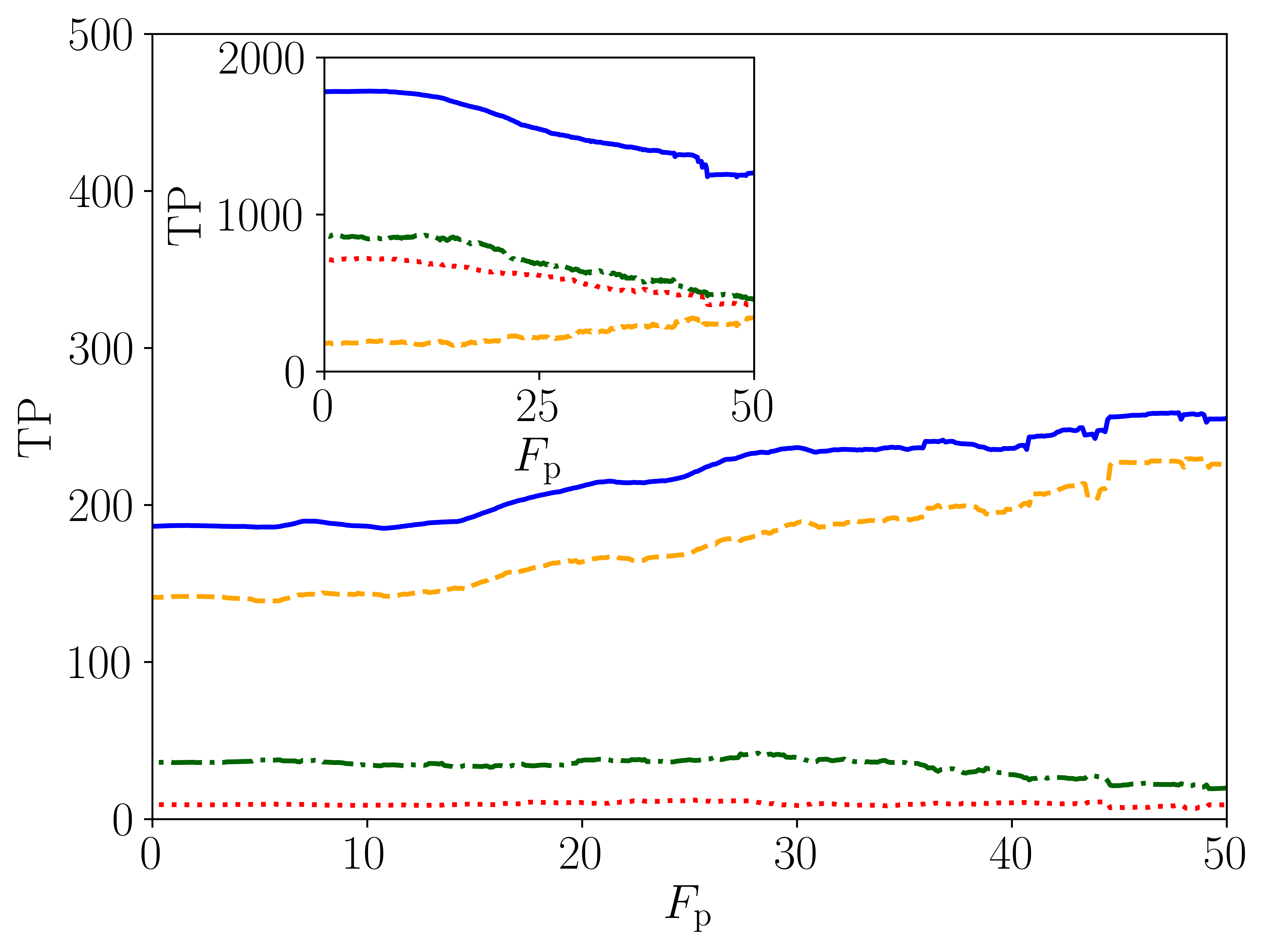}
   \caption{Total persistence, $\tp$ for $\beta_0$ structures (clusters) (main figure) and for 
   $\beta_1$ structures (loops) (inset). The \textcolor{blue}{total value (solid blue line)} of $\tp$ 
   is shown, as well as the values of $\tp$ for the birth coordinate in \textcolor{orange}{strong range (dashed orange line) ($> 2.5 $)},
   \textcolor{green}{medium range (dash-dotted green line) $[1.0~-~2.5]$} and \textcolor{red}{weak range (dotted red line) $[0.0~-~1.0]$}. 
   }
  \label{fig:TP}
\end{figure}

The concept of clusters and loops is useful since these quantities provide an initial insight, and also since they
can be related to the commonly considered `force chains' and `cycles'.  However, this concept suffers
from a significant deficiency, and that is the threshold dependence.  This deficiency is removed by the total 
persistence, which essentially merges the complete information (for all thresholds) into a single number, as 
discussed in Sec.~\ref{sec:PH}.  Figure~\ref{fig:TP} shows the total persistence, $\tp$ for $\beta_{\rm 0}$ 
(and for $\beta_{\rm 1}$ in the inset) structures.  The $\tp$ for $\beta_0$ structures (clusters) increases 
continuously with the pullout force, showing again that restructuring of the force network takes 
place (even when rescaling by the average time-dependent normal force).  We could now ask whether $\tp$ increases across the whole range of the interaction forces, or whether the increase is concentrated in a particular range.  Figure~\ref{fig:TP} shows that the latter is correct: the values of $\tp$ for the $\beta_0$ structures 
characterized by the force significantly
larger than the average force, $F_{\rm n,a}$, increase strongly, while the $\tp$ for the structures
born at the medium or weak level remain essentially the same.  The inset of Fig.~\ref{fig:TP} shows that the 
trend for $\beta_1$ structures (loops) is different, since $\tp$ decreases as $F_{\rm p}$ increases; we discuss
this behavior further below.  

\begin{figure}[ht]
\centering
  \includegraphics[width=0.475\textwidth]{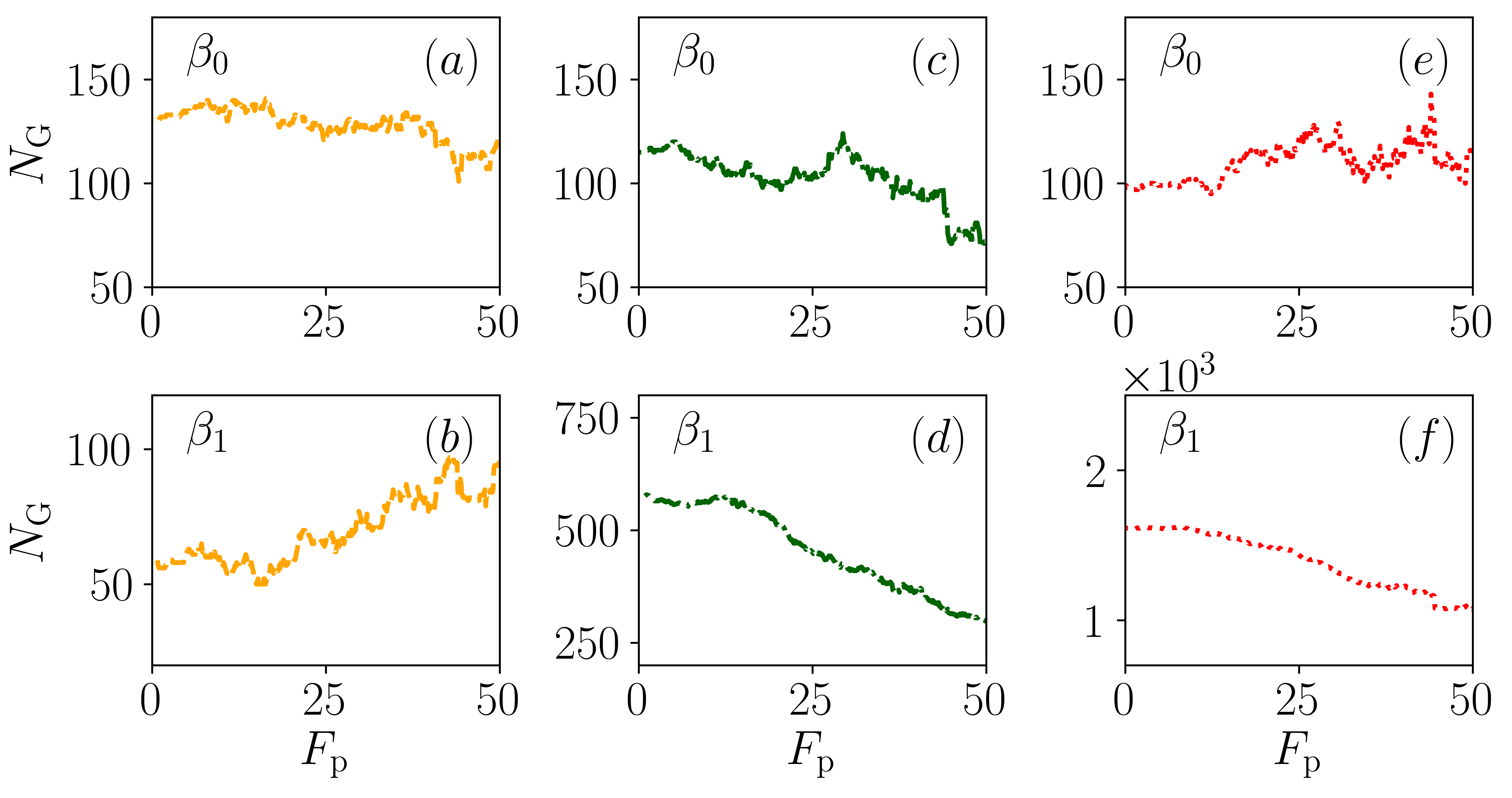}
   \caption{Number of generators (points in the $\pd$'s) using the same line and color convention as in Fig.~\ref{fig:TP}:
   \textcolor{orange}{strong (a, b)}, \textcolor{green}{medium (c, d)}, and \textcolor{red}{weak (e, f)}.
      } 
  \label{fig:generators}
\end{figure}

One could ask what is the nature of the changes in the force networks that leads to such a 
strong increase of $\tp$ for the $\beta_0$ structures born at a large force level: to use a landscape analogy, 
one could ask whether the hills become higher, or there are more hills as $F_{\rm p}$ increases (recall that $\tp$ essentially 
combines the information about lifespans, $\mathcal{L}$, of the generators in $\pd$'s, and the number of generators).
To answer this question, Fig.~\ref{fig:generators} shows the number of generators.  
Most interestingly, we observe that the number of $\beta_{\rm 0}$ 
generators is essentially constant, saying that the reason for the increase of $\tp$ for strong forces is 
an increased nonuniformity of the interaction field: the strong generators are born at even higher forces. 
Regarding the loops, Fig.~\ref{fig:generators}b), d), f) show an increase of the number of loops in 
the strong region, on the expense of the medium ones, consistently with the results for $\tp$ in Fig.~\ref{fig:TP} (inset); however
overall number of generators decreases, leading do a decrease of total $\tp$.  Supplementary Material~\cite{supp_mat_PD}
provide more detailed insight regarding the force network evolution prior to failure.

\begin{figure}[ht]
\centering
  \includegraphics[width=0.475\textwidth]{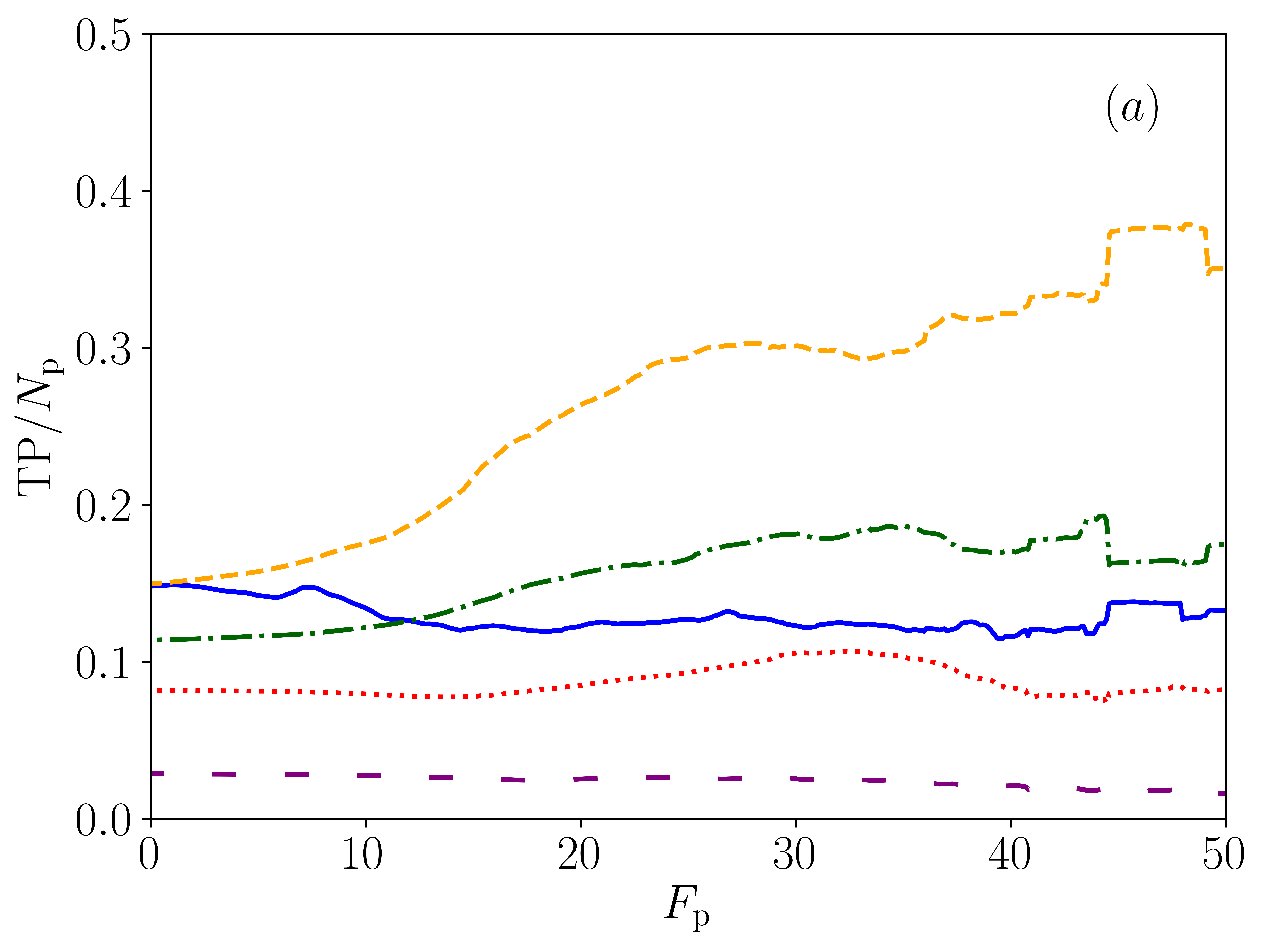}
   \includegraphics[width=0.475\textwidth]{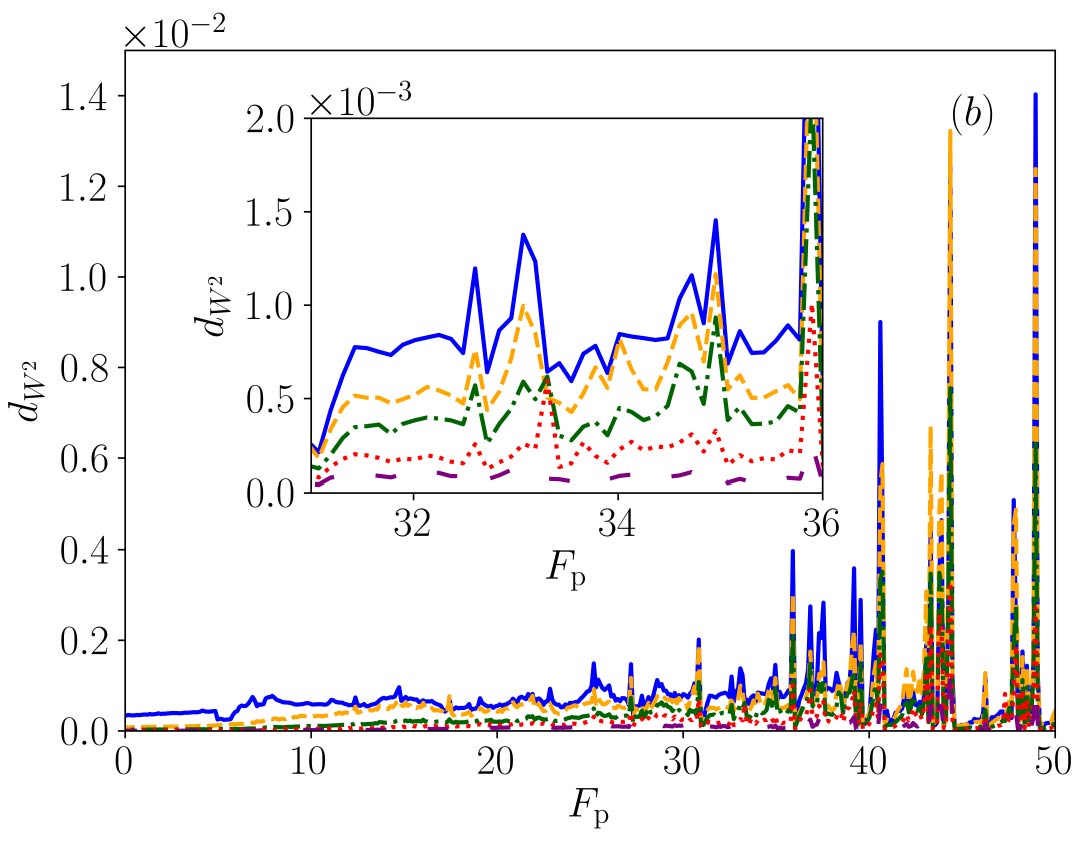}
  \caption{(a) Total persistence, $\tp$ for $\beta_0$ structures (clusters), computed separately for the five parts of the system, 
  for \textcolor{blue}{$0 < z < 1$ (solid blue line)}, \textcolor{orange}{$1 < z < 1.72$ (dashed orange line)}, 
  \textcolor{green}{$1.72 <z < 2.44$ (dash-dotted green line)}, 
  \textcolor{red}{$2.44 <z< 3.16$ (dotted red line)}, and 
  \textcolor{purple}{$ 3.16 <z < 4.18$ (loosely dashed purple line)}. 
  (b) Distance $d_{W^2}$ for $\beta_0$ structures for the same five regions as in (a), using the same convention 
  for lines patterns and colors.  The inset 
  expands a part of the considered range of the pulling force, $F_{\rm p}$.
   }
  \label{fig:parts}
\end{figure}

Another question to ask is whether the force network changes everywhere as the pullout
force increases, or whether the changes are localized to a specific 
part of the system.  To answer this question, we
compute $\tp$ separately for five different parts of the system,  according to the distance from $z=0$ (bottom of the column).  
Figure~\ref{fig:parts}a) show the results; for simplicity, here we consider
$\beta_0$ $\tp$ only, and also normalize $\tp$ by the number of particles, $N_{\rm p}$,  in each of the separate regions.  
We see that the changes of $\tp$ are concentrated in
the parts of the system that are close to the intruder, with the most pronounced changes in the layer just
above the intruder itself.

So far, we have focused on the analysis of system properties at a specified time instant, without
attempting to discuss temporal evolution of the force network.  However, temporal evolution is very 
much of interest as well, in particular since the animations in Supplementary
Materials~\cite{supp_mat_DEM,supp_mat_PD} suggest that time dependency of force networks is 
far from simple.  To gain at least basic insight into this evolution, 
we use the concept of distance between $\pd$s, discussed in Sec.~\ref{sec:PH}, see Eq.~(\ref{eq:dist}),
normalized by the number of particles in each region, $N_{\rm p}$.  
Figure~\ref{fig:parts}b) shows the results, again for the five regions mentioned above.
For simplicity, we consider only $d_{W^2}$ distance.   Figure~\ref{fig:parts}b) shows the 
localized (both in space
and time) changes of the distance, which are particularly prominent in the region close to the intruder
itself.  The inset of Fig.~\ref{fig:parts}b) shows expanded view, illustrating sudden changes of the
considered distance measure.  
These results show clearly that the force network starts evolving while the 
particles (and the contact network) are essentially static, and that the changes of the force 
network are not continuous, but consist 
of short bursts of activity, which become more prominent as the system approaches failure.  

As mentioned at the beginning of Sec.~\ref{sec:networks}, we focus on the normal forces 
only in this section.  While we found that the tangential forces provided relevant information 
about the mechanism of failure as well, the measures obtained by computing persistence and 
derived quantities turn out to be qualitatively similar to the ones obtained based on the normal 
forces only, so for brevity we do not discuss them further.   

\section{Conclusions}\label{sec:conclusions}

DEM simulations of the pullout of an intruder buried in confined granular media show 
that the force required to fail the material varies both with the height of the 
granular material and with the column diameter.  Frictional effects are found to be important
for large material heights only, showing that in hydrostatic regime (for small filling 
heights), the frictional effects are not relevant.  While for smaller pullout forces the 
force exerted by the granular particles on the sidewall remains small, for larger pullout
forces, these two forces change at the same pace, illustrating the
role that the sidewalls and particle-wall interactions play in the failure process.  

Analysis of the interparticle forces, using both classical and novel methods based on persistent homology, uncovers
the details of the failure process.  While the intruder and the granular particles are 
essentially static before failure, and characterized by a (static) contact network, 
the (weighted) force network that includes the information about the particle
contacts goes through significant changes. The tangential forces between the 
particles increase as the applied force becomes larger, leading to a larger number of 
sliding contacts. The normal force between the particles evolves as well, and the 
force network based on the normal force becomes more structured, involving 
larger number of loops and stronger forces between the particles, even when normalized
by the mean.  The failure itself occurs through a culmination of temporally intermittent
changes that take place predominantly in the physical proximity to the intruder. Furthermore, 
the changes in the force network are focused in particular on the strong
forces when considering clusters of contacts, while the loops that form in a network are
evolving at all considered levels of the interaction strength.  

By combining classical and newly developed approaches for describing both statics and dynamics
of a granular system, we have discussed in this work precise and objective measures that can
be used to describe a failure process in the considered system.  It remains to be seen to which 
degree the presented findings apply to other systems close to failure.  We hope that both the 
methods and the results presented in the current paper will provide guidance for future research 
in this direction. 

\section*{Conflicts of interest}
There are no conflicts to declare.

\section*{Acknowledgements}
Shah and Jalali would like to acknowledge the financial support provided by the Academy of Finland under Grant No. 311138. Cheng and Kondic acknowledge support by the ARO Grant No. W911NF1810184.






%

\end{document}